\documentclass[groupedaddress,superscriptaddress,aps,showpacs,amsmath,floatfix, prl,twocolumn,a4]{revtex4-1}

\usepackage{graphicx,float}
\usepackage{subfigure}
\usepackage{dcolumn}
\usepackage{bm}
\usepackage{mathrsfs}
\usepackage{txfonts}
\usepackage{CJK}
\usepackage[amssymb]{SIunits}
\usepackage{lipsum}
\usepackage{color}
\usepackage{textcomp}
\usepackage{inputenc}

 \newenvironment{SChinese}{%
  \CJKfamily{gbsn}%
 \CJKtilde
  \CJKnospace}{}

\allowdisplaybreaks[2]

\begin{document}

\begin{CJK}{UTF8}{} 
\begin{SChinese}

\title{Hybrid quantum gate and entanglement between a ``stationary'' photonic qubit and a flying optical state via a giant cross Kerr nonlinear effect}

 \author{Keyu Xia (夏可宇)}  %
 \email{keyu.xia@mq.edu.au}
 \affiliation{ARC Centre of Excellent for Engineered Quantum Systems, Department of Physics and Astronomy, Macquarie University, NSW 2109, Australia}

 \author{Jason Twamley}
 \email{jason.twamley@mq.edu.au}
 \affiliation{ARC Centre of Excellent for Engineered Quantum Systems, Department of Physics and Astronomy, Macquarie University, NSW 2109, Australia}

\date{\today}

\begin{abstract}
  Quantum information processing with hybrid protocols making use of discrete- and continuous-variable currently attracts of great interest because of its promising applications in scalable quantum computer and distant quantum network. By inducing a giant cross-Kerr nonlinearity between two cavities, we propose a general protocol for hybrid quantum gate and quantum entanglement with high fidelity between a stationary, discrete photonic qubit and a flying photonic state. Interestingly, our protocol can be used to conduct a controlled-Z quantum gate between a stationary microwave photon stored in a slowly-decaying microwave cavity and a flying optical photon, and therefore enable to build quantum network for distant superconducting quantum circuits.
\end{abstract}


\maketitle

\end{SChinese}
 \end{CJK}
 
 \emph{Introduction.\texttwelveudash} Hybrid quantum information processing (QIP) with the stationary discrete qubit and a flying photon can enable scalable quantum computation, distributed quantum information processing and distant quantum network \cite{NatPhys.11.713,Nature.484.195}. The hybrid entanglement, as the heart of hybrid QIP, between a discrete single-photon qubit and a coherent-state field has only been demonstrated very recently by using heralded frequency conversion \cite{NatPhoton.8.564,*NatPhoton.8.570}. 
 An entangling hybrid quantum gate being able to control the quantum state of a flying photon with a stationary discrete qubit in the quantum regime is highly demanded because it allows the linking of distant computation node to build a quantum network. The development of such quantum gate is ``a pressing challenge'' \cite{Science.339.1174}. The hybrid quantum gate between a trapped atom and a flying photon has been reported in waveguide quantum electrodynamics (wQED) systems \cite{PhysRevLett.111.090502,PhysRevLett.109.160504} and cavity quantum electrodynamics (cQED) systems \cite{Nature.484.195, Nature.508.237,SciRepsGongwei}. However, a protocol for the hybrid quantum gate between a discrete photonic qubit and a flying photon is yet to be proposed. 
 
 Superconducting quantum circuits working in the microwave regime are building blocks for the most promising quantum computer \cite{Science.339.1169} but can only communicate via their mutual interaction with microwave (mw) photons. Microwave photons, unlike optical photons, cannot be efficiently transmitted over long distances. To circumvent this challenge, schemes converting the quantum information between the mw and optical (mw-o) regimes have been proposed by using optomechanics \cite{PhysRevLett.109.130503,*PhysRevLett.108.153603, *PhysRevLett.108.153604, *SciRep.4.5571} or spins \cite{PhysRevLett.113.203601,*PhysRevLett.113.063603,*PhysRevA.91.042307}. The frequency conversion of classical photonic information with mechanical oscillator has been realized \cite{NatPhys.9.712,*NatPhys.10.321}, but its quantum counterpart is lack. Very recently, wQED systems are exploited to create hybrid entanglement between a superconducting qubit and a flying photon  \cite{arXiv.1608.05135,*arXiv.1607.06271}. However, hybrid quantum gate bridging the mw regime and the optical regime is lack of protocol.
 
 The controlled-Z quantum gate is one of universal entangling quantum gate \cite{Nature.508.237,PhysRevLett.114.173603} and can be easily transfered to a controlled-NOT quantum gate \cite{Nature.508.237}. It has been supposed to be realized in a cross-Kerr nonlinear optical medium. The giant optical Kerr and cross Kerr nonlinearity has been predicted theoretically in N-type atoms \cite{OptLett.21.1936,*PhysRevLett.103.150503,*NatPhys.2.849,*PhysRevLett.79.1467,*PhysRevLett.82.4611,*JOB.1.490} and been observed in many experiments \cite{NatPhys.11.905,*NatPhoton.6.93,*PhysRevLett.91.093601,*PhysRevA.84.053820,*PhysRevA.83.041804,*JOpt.17.045505}. The cross Kerr nonlinearity has been widely considered as a toolkit for the phase-flip quantum gate between two flying photons for decades \cite{Nature.396.537,*PhysRevLett.84.1419,*PhysRevA.61.011801}, but its validity for QIP of continuous variables is now questioned at the fundamental level \cite{PhysRevA.73.062305,PhysRevA.81.043823}. In this letter, with a giant cross Kerr nonlinearity between a ``good'' cavity (g-cavity) and an one-side optical ``bad'' cavity (b-cavity) created by an ensemble of N-type atoms, we present a protocol for a hybrid controlled-Z quantum gate between a discrete photon stored in the slowly-decaying g-cavity and a flying photon encoded in polarization. The phase of flying photonic state reflected off the b-cavity is conditionally controlled by the quantum state of the g-cavity due to the giant cross Kerr nonlinearity. Interestingly, our protocol can conduct a hybrid quantum gate between a discrete mw photon and a flying optical photon and therefore allows the linking of distant superconducting quantum circuits. It can also generate the hybrid entanglement between a ``stationary'' discrete photonic qubit and a coherent-state field.

\begin{figure}
  \centering
  \includegraphics[width=1\linewidth]{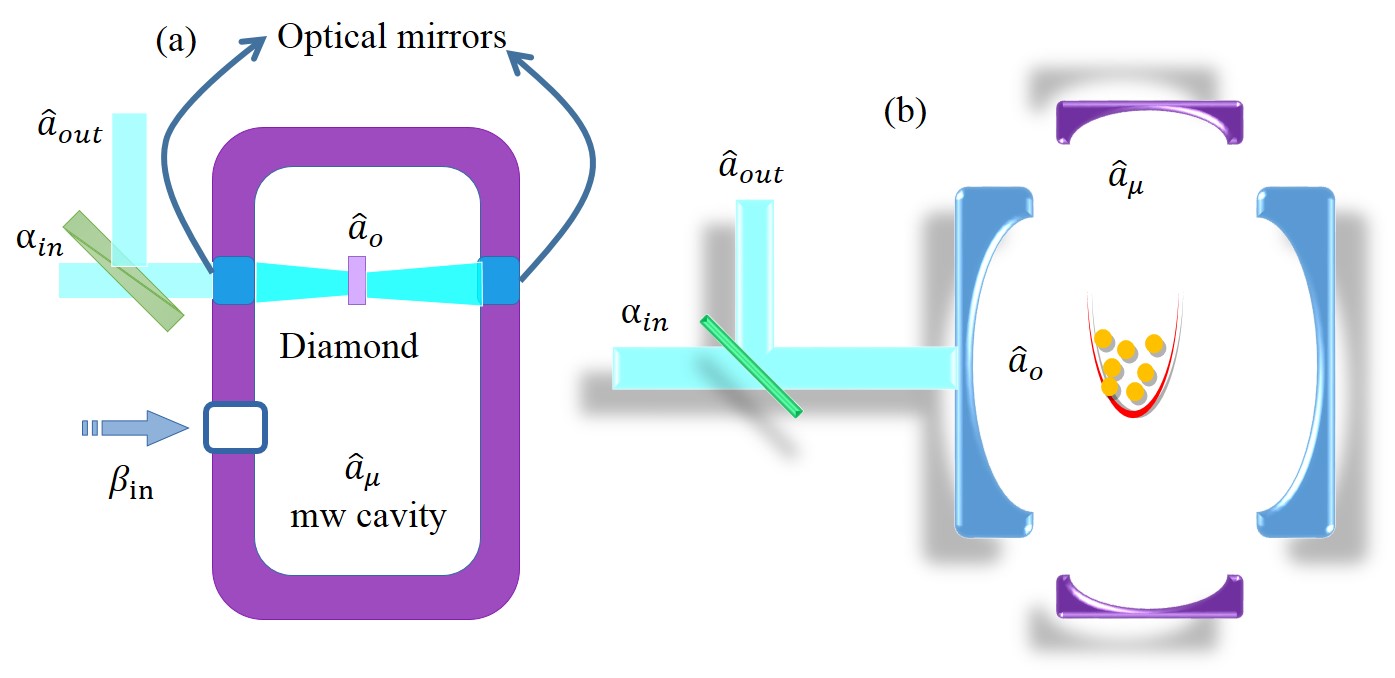}\\
  \caption{(Color online) Schematic diagram of setup controlling the reflection ($\hat{a}_\text{out}$) of an optical input $\alpha_\text{in}$ with another cavity mode $\hat{a}_\mu$. (a) An ensemble of NV centers in nanodiamond couples simultaneously to a slowly-decaying three-dimension (3D) mw cavity and a fast-decaying optical Fabry-Perot cavity. (b) A cloud of N-type cold atoms interacts with two optical Fabry-Perot cavities. The ensemble of atom/NV centers induces a giant cross Kerr nonlinearity between two cavities.}\label{fig:Implement}
 \end{figure}

 \emph{System.\texttwelveudash} Now we begin the discussion of our idea and model with the description of our setup depicted in Fig. \ref{fig:Implement}. 
 We present a general protocol for a hybrid controlled-Z quantum gate between a flying optical photon reflected off a ``bad'' optical cavity (cyan) and a ``stationary'' photon stored in a ``good'' cavity (purple) with a high quality (Q). Figure \ref{fig:Implement}(a) shows the setup for the mw-o controlled-Z quantum gate between a ``stationary'' mw photon stored in a 3D mw cavity and a flying optical photon. To do this, we make two tiny holes on the metal wall of mw cavity and put two highly reflective mirrors to form a Fabry-Perot cavity.  An ensemble of NV centers in nanodiamond is inserted in the common region of the mw cavity and the Fabry-Perot optical cavity. This optical cavity supports two degenerate modes with left-circular(or $\sigma_-$) polarization (LCP) and right-circular(or $\sigma_+$) polarization (RCP). The mw cavity mode couples to the transition of $|^3{\bf A}_2,m_s =0\rangle \leftrightarrow |^3{\bf A}_2,m_s=1\rangle$. It can be quickly initialized with a superconducting qubit \cite{Science.342.607}. The RCP(LCP) mode couples the transition between $|^3{\bf A}_2,m_s=-1\rangle$ ($|^3{\bf A}_2,m_s=1\rangle$) and the optical excited state, $|^3{\bf E},S_z\rangle$. A coherent laser beam drives the transition between $|^3{\bf A}_2,m_s=0\rangle \leftrightarrow |^3{\bf E},S_z\rangle$.
 Figure \ref{fig:Implement}(b) illustrates an optical-optical(o-o) version. Instead, the mw cavity is replaced by a slowly-decaying optical cavity with high Q (purple). The N-type atom can be a cloud of cold Rubidium (Rb) or Cesium (Cs) atoms. This ``good'' cavity can be initialized with another cloud of cold atom, which is separate from the b-cavity. 
 
 \emph{Model.\texttwelveudash} Without loss of generality, we next describe the general model for the two setups mentioned above.
 We assume that the g-cavity has a resonance frequency of $\omega_\mu$, an intrinsic loss of $\kappa_{i\mu}$ and an external coupling rate $\kappa_{e\mu}$ yielding a total decay rate of $\kappa_\mu=\kappa_{e\mu}+\kappa_{i\mu}$. We take $\kappa_{e\mu}=\kappa_{i\mu}$. As an example, we take the decay rate $\kappa_\mu/ 2\pi \sim 10~\text{\kilo\hertz}$, which is available for either a 3D mw cavity \cite{Science.342.607} or an optical cavity \cite{Nature.529.505,PhysScrip.T76.127,Science.333.1266} using the existing experimental technology.
 The flying optical photons enter and are reflected off the optical b-cavity through the same (left) mirror at a rate of $\kappa_\text{eo}$. We assume that this cavity has an intrinsic loss rate of $\kappa_\text{io}$ that the total loss is $\kappa_\text{o}=\kappa_\text{io}+\kappa_\text{eo}$. It is reasonable to assume that $\kappa_\text{eo} \gg \kappa_\text{io}$ and $\kappa_\text{o} \gg \kappa_\mu$. We take $\kappa_\text{o}/2\pi = 10~\text{\mega\hertz}$ to neglect the loss of photon in the g-cavity. The ensemble of N-type atoms simultaneously couples efficiently to both the R-polarized mode of the b-cavity ($a_\text{o}$) and the g-cavity ($a_\mu$). Due to the large detuning of the L-polarized mode of the g-cavity from the atomic transition of $|4\rangle \leftrightarrow |3\rangle$, it separates from the atoms and is reflected by an empty cavity. In the following discussion we focus on the interaction of this R-polarized cavity mode and the g-cavity mode. 
 The N-type atoms creates a large photon-photon interaction $H_\text{I}=\hbar \eta a^\dag_\text{o} a_\text{o} a^\dag_\mu a_\mu$ between these two cavities, where the interaction strength $\eta$ is determined by the cross Kerr nonlinear susceptibility $\chi^{(3)}$ to be evaluated later. When $\eta \gg \kappa_\text{o}$, we can control the reflectivity of the flying photon off the b-cavity with the quantum state of photon stored in the g-cavity.
 \begin{figure}
  \centering
  \includegraphics[width=0.66\linewidth]{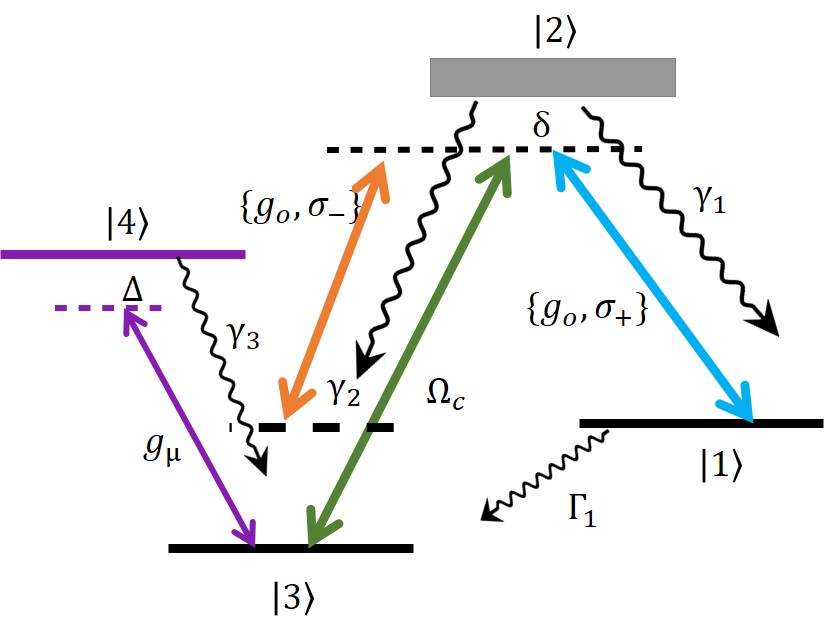}\\
  \caption{(Color online) Level diagram describing the interaction between mw ($g_\mu$) and photons ($g_\text{o}$) and an ensemble of NV centers in the nanodiamond. A classical laser field drives the transition between states $|2\rangle$ and $|3\rangle$ with a rate of $\Omega_c$.}\label{fig:level}
 \end{figure}
 
 Next we go to derive the coupling strength $\eta$.
 The interaction of cavity modes and the atoms is shown in Fig. \ref{fig:level} (a). The N-type atom has four levels, $|j\rangle$ with $j\in\{1,2,3,4\}$, which correspond to the states $\{|^3{\bf A}_2, m_s=-1\rangle,|^3{\bf E},S_z\rangle, |^3{\bf A}_2, m_s=0\rangle, |^3{\bf A}_2, m_s=1\rangle\}$ for NV centers in nanodiamond or $\{|6S_{1/2},3,3\rangle,|6P_{3/2},4,4\rangle, |6S_{1/2},4,4\rangle, |6P_{3/2},5,5\rangle \}$ for Cs atoms \cite{Science.341.768}. The state $|2\rangle$ decays to the states $|1\rangle$ and $|3\rangle$ with rates $\gamma_1$ and $\gamma_2$, respectively, while the state $|1\rangle$ and $|4\rangle$ decay to $|3\rangle$ with rates $\Gamma_1$ and $\gamma_3$. The b-cavity (g-cavity) mode couples the transition $|2\rangle \leftrightarrow |1\rangle$ ($|4\rangle \leftrightarrow |3\rangle$)  with a detuning $\delta$ ($\Delta$) and at a rate $g_\text{o}$ ($g_\mu$). 
 A classical laser field $\Omega_c$ drives the transition $|2\rangle \leftrightarrow |3\rangle$ with a detuning $\delta$. Under the condition of $\Omega_c \gg g_\text{o}$, the atoms are dominantly trapped in $|1\rangle$. Then the Hamiltonian describing the interaction between the atoms and the cavities takes the form
 \begin{equation} \label{eq:Hc}
  \begin{split}
   H_c = & \Delta \sigma_{44} + \delta \sigma_{22} + \Omega_c (\sigma_{23} + \sigma_{32}) \;\\
   & + g_\mu (\hat{a}_\mu^\dag \sigma_{34} + \sigma_{43} \hat{a}_\mu) + g_\text{o} (\hat{a}_\text{o}^\dag \sigma_{12} + \sigma_{21} \hat{a}_\text{o}) \;.
  \end{split}
 \end{equation}
 Applying the perturbation theory and adiabatically eliminating the atoms, we obtain a rough estimation of the cross third-order susceptibility given by
 \begin{equation} \label{eq:XKerr}
    \chi_\text{cross}^{(3)} \approx - \frac{2N_a d^2_{21} d^2_{43}}{\hbar^3 \varepsilon_0 V_a} \frac{1}{(\Delta - i \Gamma_{1}-i\gamma_3)(\Omega_c^2 + g_\mu^2 \langle a^\dag_\mu a_\mu\rangle)} \;,
 \end{equation}
 where $d_{21}$ ($d_{43}$) is the dipole moment of optical (mw) transition of $|2\rangle \leftrightarrow |1\rangle$ ($|4\rangle \leftrightarrow |3\rangle$), $\varepsilon_0$ is the vacuum permittivity and $V_a$ is the volume of the nanodiamond or the whole atomic cloud.
The photon-photon interaction strength can be obtained from $\Re[\chi_\text{cross}^{(3)}]$,
 \begin{equation} \label{eq:eta}
  \eta \approx - \frac{ 2N_a g_\text{o}^2g_\mu^2 \Delta}{[\Delta^2  + (\Gamma_1+\gamma_{3})^2] (\Omega_c^2 + g_\mu^2 \langle n_\mu\rangle)} \;.
 \end{equation}
 The driving of the cavities can be described by $H_\text{Dr}=i\sqrt{2\kappa_{\text{e}o}}(\alpha_\text{in} \hat{a}_\text{o}^\dag - \alpha_\text{in}^* \hat{a}_\text{o}) + i\sqrt{2\kappa_{\text{e}\mu}}(\beta_\text{in} \hat{a}_\mu^\dag - \beta_\text{in}^* \hat{a}_\mu)$, where $\alpha_\text{in}$ and $\beta_\text{in}$ are the input driving the optical cavity and mw cavity, respectively. 
 Thus we have the effective Hamiltonian for the cavity modes with the optical external input centered at frequency $\omega_\text{in}$ only 
 \begin{equation} \label{eq:Heff}
 \begin{split}
  H_\text{eff} = & (\Delta_\text{in}-i\kappa_\text{o}) \hat{a}_\text{o}^\dag \hat{a}_\text{o} -i\kappa_\mu \hat{a}_\mu^\dag \hat{a}_\mu + \eta\hat{a}_\mu^\dag \hat{a}_\mu \hat{a}_\text{o}^\dag \hat{a}_\text{o} \\
   & + i2\kappa_{\text{e}o}[\hat{a}_\text{in}(\omega_\text{in}) \hat{a}_\text{o}^\dag - \hat{a}_\text{in}^\dag(\omega_\text{in}) \hat{a}_\text{o}]\;,
  \end{split}
 \end{equation}
 where $\Delta_\text{in}= \omega_\text{o} - \omega_\text{in}$, and $[\hat{a}_\text{in}(\omega^\prime_\text{in}),\hat{a}_\text{in}^\dag(\omega_\text{in})]=\delta(\omega^\prime_\text{in}-\omega_\text{in})$. For our purpose, here external driving, $\beta_\text{in}$, of the g-cavity is tuned off. In contrast to the RCP input, the LCP incident photon is scattered by an empty cavity, i.e. $\eta=0$. 
 
 Now we can discuss our hybrid quantum gate based on the effective Hamiltonian $H_\text{eff}$. Using the input-output relation for the reflection of optical cavity \cite{PhysRevA.30.1386,PhysRevA.31.3761}, the amplitude of reflection the takes
 \begin{equation} \label{eq:reflection}
  r_\pm(\omega_\text{in},\hat{N}_\mu) = \frac{-i(\omega_\text{o}-\omega_\text{in}  + \eta_\pm \hat{N}_\mu) + \kappa_\text{eo}-\kappa_\text{oi}}{i(\omega_\text{o}-\omega_\text{in}  + \eta_\pm \hat{N}_\mu) + \kappa_\text{eo}+\kappa_\text{oi}} \;,
 \end{equation}
 with $\hat{N}_\mu=a^\dag_\mu a_\mu$, $\eta_-=0$ for the LCP incident and $\eta_+=\eta$ for the RCP one. The annihilation operator of the reflected field is given by $\hat{a}_s=r_\pm(\omega_\text{in},\hat{N}_\mu) \hat{a}_\text{in}$. 
 If $\omega_\text{o}-\omega_\text{in}=-\eta$ and $|\eta| \gg \kappa_\text{o}$, then we have $r_+(\omega_\text{in},\hat{N}_\mu) \approx -e^{i\pi \hat{N}_\mu}$ and $r_-(\omega_\text{in}) \approx -1$. It means the LCP incident photon is always reflected and suffers a $\pi$ phase shift, whereas the reflection of the RCP input is controlled by the quantum state of the g-cavity. If a continuous-wave (cw) RCP photon enters the b-cavity, it is reflected off the cavity without phase shift when $\langle N_\mu \rangle=1$. When $\langle N_\mu \rangle=0$, the RCP input photon is directly reflected off the cavity with a $\pi$ phase shift.
 As a result, a cw RCP(LCP) incident photon in state $|\omega\rangle = \hat{a}_\text{in}(\omega)|\varnothing\rangle$, where $|\varnothing\rangle$ is the vacuum state in open space, after reflected off the g-cavity, becomes $r_+(\omega, \hat{N}_\mu)|\omega\rangle$ ($r_-(\omega,\hat{N}_\mu)|\omega\rangle$). The b-cavity mode $\hat{a}_\text{o}$ is related to the incident mode $\hat{a}_\text{in}$ via
 \begin{equation} \label{eq:ao}
  \hat{a}_\text{o}(\omega_\text{in},\hat{N}_\mu) = \frac{2\kappa_\text{eo}}{i(\omega_\text{o}-\omega_\text{in}  + \eta_\pm \hat{N}_\mu) + \kappa_\text{eo}+\kappa_\text{oi}} \hat{a}_\text{in}(\omega_\text{in})\;.
 \end{equation}
 Here we choose $\hat{a}_\text{o}$ and $\hat{a}_\text{in}$ to be dimensionless operators.
 The photon-photon interaction $H_\text{I}$ also induces phase shift $\theta$ between the states $|0\rangle$ and $|1\rangle$ of the g-cavity mode, taking $\theta = \eta_\pm \int \langle \hat{a}_\text{o}^\dag(\tau) \hat{a}_\text{o}(\tau) \rangle d\tau=\eta_\pm \int \langle \hat{a}_\text{o}^\dag(\omega) \hat{a}_\text{o}(\omega) \rangle d\omega$.

 Next we go to evaluate the performance of the hybrid quantum gate and the generated hybrid entanglement for a Gaussian-pulsed input $s(t)=\frac{1}{\sqrt{\sqrt{\pi} \sigma_T}} e^{-t^2/2\sigma_T}$ on the input mirror of the b-cavity with a duration of $\sigma_T=5 \kappa_\text{o}$ corresponding to a spectrum $f(\omega)= \frac{1}{\sqrt{\sqrt{\pi}\sigma_\omega}}e^{-(\omega-\omega_\text{in})^2/2\sigma_\omega^2}$ and $\sigma_\omega=1/\sigma_T$. 
 
 \emph{Controlled-Z quantum gate for polarization flying qubit.\texttwelveudash}
 Here we evaluate the hybrid quantum gate which can impose a $\pi$ phase flip on a flying qubit encoded in polarization with the quantum state of the g-cavity mode. We use the notations $|R(\omega)\rangle=\hat{a}^\dag_R(\omega)|\varnothing\rangle$ and $|L(\omega)\rangle=\hat{a}^\dag_L(\omega)|\varnothing\rangle$ for indicating an incident photon with frequency $\omega$ and RCP or LCP, respectively. The Gaussian input photon with spectrum $f(\omega)$ encoded in the RCP(LCP) can be denoted as $|R\rangle = \int f(\omega) |R(\omega)\rangle d\omega$ ($|L\rangle = \int f(\omega) |L(\omega)\rangle d\omega$) \cite{PhysRevLett.111.090502,PhysRevLett.114.173603}. Initially, the g-cavity is prepared in $\alpha_\mu|0\rangle + \beta_\mu |1\rangle$ and the incident photon in $\alpha_\text{o}|L\rangle + \beta_\text{o} |R\rangle$. Ideally, the controlled-Z gate converts the initial states in the way of $|0,L\rangle \rightarrow -|0,L\rangle, |0,R\rangle \rightarrow -|0,R\rangle,|1,L\rangle \rightarrow -|1,L\rangle$ and $|1,R\rangle \rightarrow |1,R\rangle$. This process defines a target unitary, $U= - |0,L\rangle \langle 0,L| - |0,R\rangle \langle 0,R|- |1,L\rangle \langle 1,L|+ |1,R\rangle \langle 1,R|$. For the incident pulse with the spectrum $f(\omega)$, our system scatters the states as $|0,L\rangle \rightarrow |0\rangle \otimes \int f(\omega) r^*_-(\omega,0)|L(\omega)\rangle d\omega, |0,R\rangle \rightarrow |0\rangle \otimes \int f(\omega) r^*_+(\omega,0)|R(\omega)\rangle d\omega, |1,L\rangle \rightarrow  |1\rangle \otimes \int f(\omega) r^*_-(\omega,0)|L(\omega)\rangle d\omega$  and $|1,R\rangle \rightarrow  e^{i\theta_\mu}|1\rangle \otimes \int f(\omega) r^*_+(\omega,1)|R(\omega)\rangle d\omega$. A quantum gate can be identified with the quantum process tomography (QPT) using a fixed basis of operators \cite{QPT1,QPT2,QPT3,QPT4}. To evaluate the quality of our controlled-Z quantum gate, we first initialized the system in a complete basis of input states $\rho_j=
 |\psi_j\rangle \langle \psi_j|$, where $|\psi_j\rangle = \{|0\rangle,|1\rangle, \frac{1}{\sqrt{2}}, \frac{1}{\sqrt{2}} (|0\rangle +i |1\rangle)\}$ for the g-cavity and 
 $|\psi_j\rangle = \{|L\rangle,|R\rangle, \frac{1}{\sqrt{2}} (|L\rangle + |R\rangle), \frac{1}{\sqrt{2}} (|L\rangle +i |R\rangle)\}$ for the flying incident photonic qubit, respectively. Then we apply the process $\mathscr{\varepsilon}$ of the ideal and proposed quantum gates to each member $\rho_j$. After that, we reconstruct the output states of the ideal gate and our proposed gate, $\rho_\text{U}$ and $\rho_\varepsilon$, using quantum state tomography. For the process, $\varepsilon$, of our proposed gate, and the target unitary, $U$, the process fidelity and the trace distance, $F_G$ and $D_G$, respectively, are $F_G=tr\left(\sqrt{\sqrt{\rho_\text{U}}\rho_\varepsilon \sqrt{\rho_\text{U}}} \right)^2$ and $D_G=\frac{1}{2}tr|\rho_\text{U}-\rho_\varepsilon|$ \cite{QPT1,QPT2}. Note that $F_G$ indicates how close to the ideal gate the proposed quantum gate can be, and $D_G$ gives the bound of the average probability of error experienced during quantum computation with the proposed gate.
 
\begin{figure}
  \centering
  \includegraphics[width=0.7\linewidth]{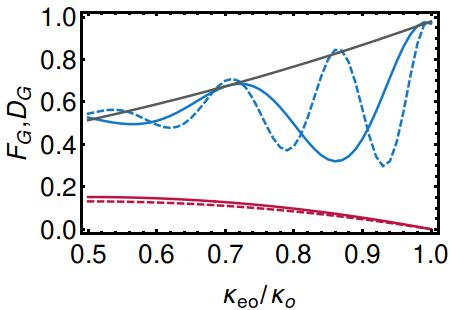} \\
  \caption{(Color online) Fidelity (blue lines) and distance (red lines) of the proposed controlled-Z quantum gate as a function of the external coupling rate. The gray line show the fidelity of the quantum gate when the backaction phase shift of the g-cavity mode is corrected. The solid (dashed) lines are for $4\eta/2\pi\kappa_\text{o}=2.2$ ($4\eta/2\pi\kappa_\text{o}=4.2$).}\label{fig:Gate}
 \end{figure}

 The photon-photon interaction can be adjusted by tuning the classical driving $\Omega_c$. We are more interested in the influence of the imperfection of the external coupling $\kappa_\text{eo}$ hard to turn, shown in Fig. \ref{fig:Gate}. The fidelity $F_G$ oscillates with an exponentially decaying amplitude as the external coupling decreases from a complete overcoupling regime, $\kappa_\text{eo}/\kappa_\text{o}=1$, to the critical coupling regime, $\kappa_\text{eo}/\kappa_\text{o}=0.5$. This oscillation is caused by the backaction phase shift $\theta$. In comparison with the case of $4\eta/2\pi=2.2$, the oscillation for a larger photon-photon interaction, $4\eta/2\pi=4.2$, is mostly twice faster. In contrast, the distance $D_G$ increases slowly as $\kappa_\text{eo}/\kappa_\text{o}$ decreasing. When $\kappa_\text{eo}/\kappa_\text{o}=0.99$, $F_D>0.96$ and $D_G\leq 0.006$. It means that our quantum gate is very close to an ideal controlled-Z quantum gate and its operation only leads to a very small error. If we can correct the backaction phase shift, $\theta$, with the initializing atoms, then the fidelity (gray line) is high and robust against the imperfection of the optimal photon-photon interaction and the external coupling.

 \emph{Hybrid entanglement and Kitten states.\texttwelveudash} 
 Here we aim to generate a target state of $|\Psi\rangle_\text{tgt}=|0_\mu,-\alpha_\text{o}\rangle + |1_\mu,\alpha_\text{o}\rangle$. Therefore, we only need a RCP incident coherent state $|\alpha_\text{o}\rangle$. For simplicity, we rewrite it as $|\Psi\rangle_\text{tgt}=|0,-\alpha\rangle + |1,\alpha\rangle$. To analysis the fidelity of the generated state $|\Psi\rangle_\text{gen}$, we express the incident field as $|\alpha\rangle = \int f(\omega) |\alpha(\omega)\rangle d\omega$, where $|\alpha(\omega)\rangle=D[\alpha,\omega]|\varnothing\rangle=e^{(\alpha \hat{a}^\dag(\omega)-\alpha^* \hat{a}(\omega))}|\varnothing\rangle$ and $\int |f(\omega)|^2 d\omega=1$. After the incident coherent-state is reflected off the b-cavity, the atoms and this b-cavity are in their ground states. Therefore, they can be traced out. Our system scattered the incident field via the b-cavity as $|0,\alpha\rangle \rightarrow |0\rangle \otimes \int f(\omega)D[r^*_+(\omega,0)\alpha,\omega] d\omega |\varnothing\rangle = |0\rangle \otimes \int f(\omega) |r^*_+(\omega,0)\alpha \rangle d\omega$ and $|1,\alpha\rangle \rightarrow e^{i\theta}|1\rangle \otimes \int f(\omega) |r^*_+(\omega,1)\alpha \rangle d\omega$. The generated state is given by $|\Psi\rangle_\text{gen}\rangle=|0\rangle \otimes \int f(\omega) |r^*_+(\omega,0)\alpha \rangle d\omega +e^{i\theta}|1\rangle \otimes \int f(\omega) |r^*_+(\omega,1)\alpha \rangle d\omega $. Its fidelity can be evaluated as $F_\text{ent} = \left|\langle \Psi|_\text{tgt} |\Psi\rangle_\text{gen}\rangle \right|^2 = \left| \int |f(\omega)|^2  [\langle-\alpha(\omega)|r^*_+(\omega,0)\alpha(\omega)\rangle +\langle\alpha(\omega)|r^*_+(\omega,1)\alpha(\omega)\rangle] d\omega \right|^2$. 
 
\begin{figure}
  \centering
  \includegraphics[width=1\linewidth]{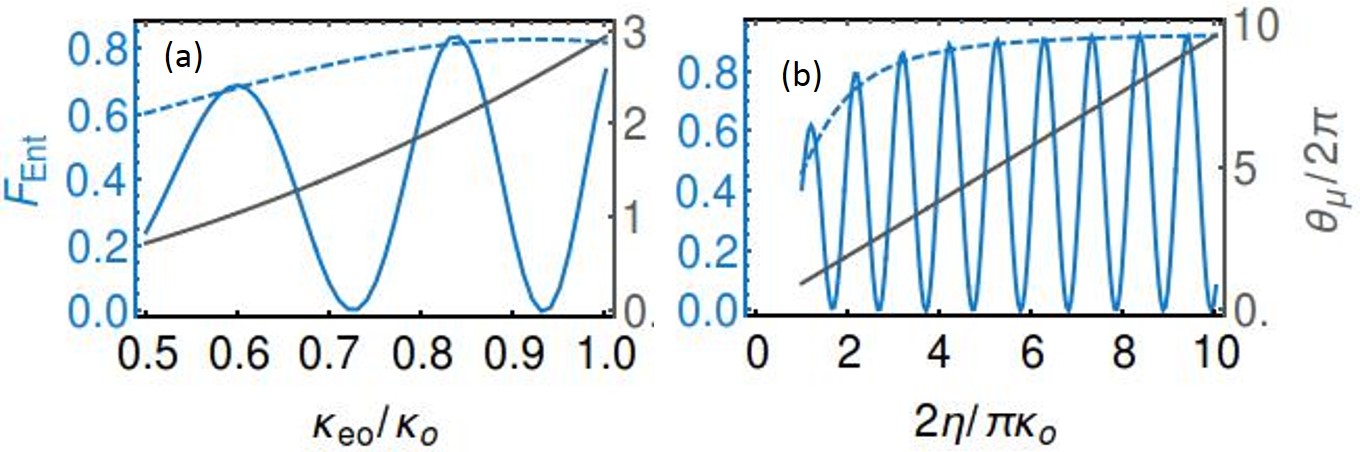} \\
  \caption{(Color online) Fidelity (blue lines) and phase shift (gray lines) of the generated hybrid entangled state as a function of the external coupling rate (a), and the photon-photon interaction strength (b) for a Gaussian-pulsed coherent-state input $|\alpha=1\rangle$. In (a), $4\eta/\kappa_\text{o}=6\pi$ and $\Delta_\text{in}=-\eta$ ; in (b) $\kappa_\text{eo}=0.99\kappa_\text{o}$ and $\Delta_\text{in}=-\eta$.}\label{fig:Ent}
 \end{figure}

 We take $\alpha=1$. The phase shift roughly approximates to $4\eta |\alpha|^2/\kappa_\text{o}$ if the bandwidth of the incident photon is much smaller than $\kappa_\text{o}$ and $\kappa_\text{eo}\approx \kappa_\text{o}$. As shown in Fig. \ref{fig:Ent}, the fidelity oscillates as the external coupling rate $\kappa_\text{eo}$ and the photon-photon interaction strength $\eta$ increase because the phase shift $\theta$ is mostly a linear function of  $\kappa_\text{eo}/\kappa_\text{o}$ and $\eta$. Figure \ref{fig:Ent}(a) shows the dependence of the fidelity on the external coupling. We take $4\eta |\alpha|^2/\kappa_\text{o}=3\times 2\pi$. When $\kappa_\text{eo}=0.83\kappa_\text{o}$, $\theta/2\pi\approx 2.03$ and the fidelity is $F_\text{Ent}\approx 0.83$. While the fidelity decreases to $F_\text{Ent}\approx 0.61$ when $\kappa_\text{eo}=0.99\kappa_\text{o}$ yielding $\theta/2\pi \approx 2.88$. The larger discrepancy of $\theta$ and $2m\pi$ when $\kappa_\text{eo}=0.99\kappa_\text{o}$ causes the decrease of the fidelity in comparison with the case of $\kappa_\text{eo}=0.83\kappa_\text{o}$. Similar influence of the phase shift $\theta$ can be seen in \ref{fig:Ent}(b) in which we take $\kappa_\text{eo}=0.99 \kappa_\text{o}$. It can be seen that the fidelity oscillates fast as a cosine function as $\theta/2\pi (\sim 4\eta|\alpha|^2/\kappa_\text{o})$ increases. By carefully choosing $4\eta|\alpha|^2/2\pi\kappa_\text{o}=\{3.19,4.21,5.24\}$ leading to $\theta/2\pi =\{3.07,4.05, 5.04\}$, we obtain $F_\text{Ent}=\{0.86,0.89,0.91\}$. Note that the optimal photon-photon interaction and the achieved fidelity are also dependent on the mean photon number, $|\alpha|^2$, of the incident field state. If we quickly correct the phase shift on the g-cavity mode by inducing a dispersive interaction with the initialing qubit/atoms after the quantum gate operation, we may achieve a high fidelity robust against the variety of the external coupling and the photon-photon interaction. Obviously, measuring the g-cavity mode in the basis of $|\pm \rangle = (|0\rangle +e^{i\theta} |1\rangle)/\sqrt{2}$ will project the generated state into one of Kitten states $(|-\alpha\rangle \pm |\alpha\rangle)/\sqrt{2}$ with the same fidelity but $50\%$ probability for each.

 \emph{Feasibility.\texttwelveudash} As mentioned in model section, the g-cavity with a decay rate of $\kappa_\mu=10^{-3}\kappa_\text{eo}$ can be considered as ``stationary'' and its loss is negligible during quantum gate operation. Therefore, the realization of the proposed quantum gate crucially relies on the available photon-photon interaction in both the mw-o and the o-o quantum gates. In both versions, we can easily engineer the system that $\Omega_c=10 g_o$. For the mw-o version, the g-cavity can be initialized with a superconducting transmon \cite{3DMWCavity}. We use an ensemble of NV centers in nanodiamond with a number density $\rho_N=5\times 10^{18}~\text{\centi\meter}^{-3}$ \cite{Nature.478.221} to induce the giant cross-Kerr nonlinearity between the mw and optical cavities. We take $\Delta/2\pi \sim 0.1~\text{\mega\hertz}\gg (\Gamma_1+\gamma_3)$ ($\Gamma_1\approx\gamma_3 \approx 2\pi \times 3 ~\text{\kilo\hertz}$ \cite{NVT2a,*NVT2b}), $g_\mu/2\pi\sim 10 \hertz$. A nanodiamond with volume of $\sim 0.7\times 10^{-6}~\text{\centi\meter}^{3}$ including $\sim 3.5\times 10^{12}$ NV centers can generate a photon-photon interaction of $4\eta/2\pi\kappa_\text{o}=2.2$. With such interacavity nonlinearity the fidelity and distance of the quantum gate can be $F_G=0.98$ and $D_G=0.006$.
 To conduct the o-o quantum gate, a cloud of cold Cs atoms is applied to generate the photon-photon interaction. A second cloud of cold atoms is applied to initialize the g-cavity. Typically, we have $\gamma_3\approx 2\pi \times 5~\text{\mega\hertz}\gg \Gamma_1$ \cite{Science.333.1266}. The single-atom coupling can be $g_\mu=2\pi \times 0.5~\text{\mega\hertz}$ or even stronger \cite{Science.333.1266}. We take $\Delta=2\pi \times 50~\text{\mega\hertz}$. Thus, A cloud of $\sim 6900$ atoms allows to obtain $F_G=0.98$ and $D_G=0.006$.
 Note that we can tune the classical driving to adjust the photon-photon interaction and then achieve the optimal fidelity. It is also noticeable that we can prepare the g-cavity in the superposition state of $(\alpha_\mu |0\rangle + \beta_\mu |N_\mu\rangle)$ with $N_\mu\gg 1$ to guarantee a high fidelity quantum gate when $\eta$ is small \cite{3DMWCavity}. 
 
 \emph{Conclusion.\texttwelveudash} With a giant intercavity cross-Kerr nonlinear interaction we have proposed a general protocol for a controlled-Z quantum gate entangling a discrete photonic qubit and a flying photon. In particular, our protocol can be applied to conduct a hybrid quantum gate between the mw regime and the optical regime: controlling the phase of the reflected flying optical qubit encoded in polarizations via the quantum state of the mw photon stored in the 3D mw cavity. Such hybrid mw-o quantum gate is an important resource for building a scalable long-distance quantum network for superconducting quantum circuits. Moreover, our protocol allows to create hybrid entanglement between a discrete photonic qubit and a flying coherent-state field, and to generate kitten state via measurements.
 
 This work was partially funded by the Australian Research Council Centre of Excellence for Engineered Quantum Systems (EQuS), Project No. CE110001013.
  

%

\end{document}